\documentclass[twocolumn,prl]{revtex4}
\usepackage{graphicx}    
\usepackage{dcolumn}     
\usepackage{bm}          
\begin{document}
\preprint{Circuit}

\title{Circuit theory of
unconventional  superconductor junctions
}
\author{
Y. Tanaka$^{1,2,3}$, Yu. V. Nazarov$^2$, and S. Kashiwaya$^{3,4}$
}%
%
\affiliation{
$^1$Department of Applied Physics,
Nagoya University, Nagoya, 464-8603, Japan \\
$^2$ Department of Nanoscience, Faculty of Applied Sciences, 
Delft University of Technology, 2628 CJ Delft The Netherlands \\
$^3$
CREST Japan Science and Technology Cooperation (JST) 464-8603
Japan \\
$^4$National Institute of Advanced Industrial Science
and Technology, Tsukuba, 305-8568, Japan
}
%
\date{\today}
\begin{abstract}
We extend the circuit theory
of superconductivity to cover transport and proximity effect
in mesoscopic systems that contain
unconventional superconductor junctions.
The approach fully accounts for
zero-energy Andreev bound states forming at the surface of unconventional
superconductors.
As a simple application,
we investigate the transport properties of  
a diffusive normal metal in series with a $d$-wave superconductor junction.
We reveal the competition between the formation of Andreev bound states and 
proximity effect, that depends on the crystal orientation of the junction interface.
\end{abstract}
\maketitle
In the last decade the mesoscopic superconducting systems
have been the subject of intensive experimental and theoretical research.
The transport in these system is essentially contributed by
the so-called Andreev reflection \cite{Andreev}, a unique process
specific for normal metal/superconductor interface.
Each electron reflected from the interface
 may transfer a charge $2e$ to the superconductor
\cite{Beenakker1Blonder,Zaitsev}.
The phase coherence between incoming electrons
and Andreev reflected holes persists in the normal metal at mesoscopic
length scale. This results in strong 
interference effects on Andreev reflection rate.
\cite{Hekking}.
The transport properties of mesoscopic $N/S$ junctions have been 
theoretically investigated
with various approaches, $e.g.$,
traditional nonequilibrium superconductivity approach \cite{LarkinVolkov},
tunneling Hamiltonian approach \cite{Hekking},
scattering formalism \cite{LambertBeenakker2}
and computer simulation \cite{Takane}.

One of the authors has proposed a generic circuit theory
of non-equilibrium superconductivity which accounts for the effects abovementioned.
The mesoscopic system is presented as a network of nodes and connectors.
A connector is characterized by a set of transmission coefficients
and can present anything from ballistic point contact to tunnel junction.
Full isotropization of electrons is assumed in the nodes.
This approach considerably simplifies  
a practical transport calculation,
numerical as well as analytical.
The circuit theory is based on conservation laws for so-called
spectral currents. These additional conservation laws
present interference of electrons and holes. The spectral currents
through each connector are functions of spectral vectors in the nodes.
There is one-to-one correspondence between spectral vectors and
currents and Keldysh Green functions in the underlying microscopic approach.
\cite{LarkinVolkov}
Kirchoff-type equations determine spectral currents and vectors in each node
and connector, and, consequently, electric current in the circuit.

Unconventional superconductors bring about very unusual interface physics.
The transport through the interface is influenced 
by  formation of Andreev bound states  (ABS) at this interface.
\cite{Buch,TK95,Kashi00}.
Those result from the interference of injected and reflected quasiparticles.
The ABS manifest themself as 
a zero-bias peak 
in tunneling conductance(ZBCP) \cite{TK95,Kashi00}. 
Indeed, ZBCP has been  reported
in various superconductors that have anisotropic pairing symmetry.
 \cite{Kashi00} 
The proper theory of transport in the presence of ABS has been formulated
\cite{TK95,Kashi00} for conditions of ballistic transport only.
This theory has to be revisited to account for diffusive transport in
the normal metal. The point is that the diffusive transport provides
an Andreev reflection mechanism for ZBCP which does not involve
any unconventional superconductivity. This mechanism may compete with
the formation of ABS.
The anomalous size dependence of transport in YBCO junctions
reported in recent experiment\cite{Hiromi}
seems to arise from this competition.
%
%
%
%
%
\par
All this has motivated us to extend the circuit theory to the systems containing
unconventional
superconductor junctions. We stress that this extension is 
by no means straightforward.
The circuit theory can not be directly applied to an unconventional superconductor
since it requires the isotropization. The latter is just incompatible
with mere existence of unconventional superconductivity. 
Fortunately, there is a way around. 
We concentrate on the matrix currents via the unconventional
superconductor junction to/from diffusive parts of the system.
If one knows the relation between these currents and the 
spectral vectors (isotropic Green functions)
in the diffusive part, one is able to use Kirchoff rules to complete
the evaluation of the matrix currents everywhere in the system.

This relation shall be derived from microscopic theory and
presents the main result of this work. We stress that applicability
of this relation is not restricted to circuit theory.
One can regard our result as a boundary condition for the traditional
Keldysh-Usadel equations of non-equilibrium superconductivity.\cite{LarkinVolkov} 
As an immediate application, we study 
a $d$-wave superconductor junction in series with normal metal. 
The resistance of the system appears to depend strongly
on the
angle $\alpha$ between the normal to the interface
and the robe direction of $d$-wave superconductor (misorientation angle).
This reveals the competition between the effect of 
ABS and proximity-induced  reflectionless tunneling.

To derive the relation between matrix current
and Green functions, we make use of the method
proposed in \cite{Nazarov2}. The method puts the older
ideas \cite{Zaitsev} to the framework of 
Landauer-B\"{u}ttiker scattering formalism.
One expresses the matrix current in a constriction 
in terms of one-dimensional
Green functions $\check g_{n,\sigma;n',\sigma'}(\epsilon;x,x')$,
where $n$,$n'$ and $\sigma ,\sigma'=\pm 1$ denote the
indicies of transport channels and the direction of motion
along $x$ axis, respectively. The "check" represents 
the Keldysh-Nambu structure. These Green functions are to be
expressed in terms of the transfer matrix that incorporates
all information about the scattering, and asymptotic Green
functions $\check G_{1,2}$ presenting boundary conditions deep 
in each side of the constriction.
The isotropization assumption requires that these $\check G$
do not depend on channel number. Under this assumption,
the current is universal depending on transmission eigenvalues
only. Although the isotropization assumption is good for conventional
superconductors and normal metals, it fails to grasp the physics
of unconventional superconductor where the Green function
essentially depends on the direction of motion and thus
on channel number.
To avoid this difficulty,
we restrict the discussion to a conventional 
model of {\it smooth interface},
assuming momentum conservation in the plane of the interface.
Within the model, the channel number eventually numbers
possible values of this in-plane momentum and 
the transfer matrix becomes block-diagonal
in the channel index.  
We thus solve  Green functions $\check g_{n,\sigma;n',\sigma'}(\epsilon,x,x')$
separately  for each channel. Asymptotic Green function
in the unconventional superconductor $\it does$ depend on the
direction of motion $\sigma$,
\begin{equation}
\check G_2;n,\sigma;n,\sigma 
= \check G^{(n)}_{2+} \frac{1-\sigma}{2} +\check G^{(n)}_{2-} \frac{1+\sigma}{2}
\end{equation}
reflecting different asymptotic conditions for
incoming ($\check G^{(n)}_{2+}$) and 
outgoing ($\check G^{(n)}_{2-}$) wave
in each channel.
The asymptotic Green function 
$\check G_1$ in normal metal is the
same for both waves and all channels (see Fig. 1).
All these matrices satisfy unitary relation 
$(\check{G}^{(n)}_{2 \pm})^{2}
=\check{G}_{1}^{2}=1$.

After some algebra we obtain the matrix current 
in the following form
\begin{equation}
\check{I}
= \frac{4e^{2}}{\rm{h} }
\sum_{m} [\check{G}_{1},\check{B}_m],
\label{main}
\end{equation}
\[
\check{B}_m
=\]
\[
\{
-\Xi_m[\check{G}_{1},\check{H}_{-}^{(m)-1}]
+ \check{H}_{-}^{-1}\check{H}_{+}^{(m)}
- \Xi_m^{2}\check{G}_{1} 
\check{H}_{-}^{(m)-1}\check{H}_{+}^{(m)}\check{G}_{1}  \}^{-1}
\]
\[ \times
[ \Xi_m(1 - \check{H}_{-}^{(m)-1} )
+ \Xi_m^{2} \check{G}_{1} \check{H}_{-}^{(m)-1}\check{H}_{+}^{(m)}
],
\]
\[
\check{H}^{(m)}_{\pm}=(\check{G}^{(m)}_{2+} \pm \check{G}^{(m)}_{2-} )/2.
\]

Here $\Xi_m \equiv T_m/(1 + \sqrt{1 -T_m})^{2}$ is related to
the transmission coefficient $T_m$ in a given channel $m$.
The above relation reduces to isotropic result of Ref. \cite{Nazarov2}
provided $\check G^{(n)}_{2+}=\check G^{(n)}_{2-} = \check{G}_{2}$. 
The above $4 \times 4$ matrix relation is the main result of 
the present work. It incorporates the most general situation   
and allows for many applications that involve unconventional
superconductors.
Below we provide a simple but extensive application example
that both illustrates circuit  theory method and demonstrates
an interesting interplay of ABS and proximity effect. 

The circuit is the one given in Fig. 1: diffusive conductor
of resistance $R_D$ in series with unconventional superconductor junction. 
We disregard
decoherence between electrons and holes in the diffusive conductor,
("leakage" current in terms of Ref.\cite{Nazarov2}),
this is justified at energies not exceeding Thouless energy
of this piece of normal metal. 
We restrict our attention to $d$-wave superconductor,
being the most practical example of the singlet unconventional
superconductor that preserves time reversal symmetry.
For simplicity, we have in mind a "two-dimensional" superconductor
made from the layers stacked in $z$-direction. 
$z$-axis lies in the plane of the interface and is normal to the
plane of Fig.1. The interface normal ($x$-axis) makes 
an angle $\alpha$ with the main crystal axis.
The propagation directions of the waves are thus in $xy$-plane
and are parameterized by the angle $\theta$ with $x$-axis. 
The angular dependence of the superconducting order parameter
is thus given by $\Delta(\theta) = \Delta_0 \cos(2(\theta - \alpha))$.
A scattering channel consists of an incoming wave in direction
$\pi-\theta$ and outgoing wave in the direction $\theta$.
The sums over channels can be reduced 
to integrals over $\theta$:
\begin{equation}
\sum_m \propto \int_{-\pi/2}^{\pi/2} d\theta \cos\theta
\end{equation}

The Green functions are fixed in "US" terminal and in "N" terminal,
the voltage $V$ is applied to "N" terminal. The Green function
in the node "DN", $\check G_1$ is not fixed and shall be determined from the 
balance of the matrix currents. There is a natural separation 
of balance equations for $2\times2$ spectral currents that
set advanced or retarded part of $\check G_1$ and for particle
current at a given energy that set the distribution function in
the node "DN".\cite{Nazarov1}

We address the balance of the spectral currents first.
The advanced $2 \times 2$ Green functions are fixed in "N"
and "US" terminals and read $\hat G_{N} = \tau_z$, 
$\hat G_{2\pm}=(\Delta_{\pm} \tau_x - i \epsilon \tau_z)/\sqrt{\Delta_\pm^2 - \epsilon^2}$,
$\bm{\tau}$ being Pauli matrices, $\Delta_{\pm} = \Delta_0 \cos(2(\theta \pm \alpha))$
being superconducting order parameters that correspond to direction
of incoming(outgoing) wave.
This suggests that the corresponding Green function
in "DN" node assumes a form $\sin\gamma\cdot \tau_x + \cos \gamma \cdot\tau_z$ 
where $\gamma$ is yet to be determined. $\gamma$ is the measure
of proximity effect. All spectral currents
are proportional to $\tau_y$. The spectral current 
$\bm{i}_{D}^{(s)}$ through diffusive
conductor is proportional to the spectral angle drop,(\cite{Nazarov1}),
the spectral current $\bm{i}_{B}^{(s)}$ via the interface is obtained from Eq.\ref{main}.
The balance equation thus reads
\begin{eqnarray}
\bm{i}_{B}^{(s)}+\bm{i}_{D}^{(s)}=0 \label{spectral_balance} \\
\bm{i}_{B}^{(s)}
= -\frac{2e^{2}}{\rm{h} }
\sum_{m} F(\gamma,\epsilon,T_m), \ \
\bm{i}_{D}^{(s)}
=\gamma/R_D \nonumber
\end{eqnarray}

Under conditions considered, the transport is determined
by energy-symmetric distribution function, that is conventionally
called $f_t$. \cite{LarkinVolkov} The balance of particle currents at each energy determines
this distribution function in "DN" node. We will assume that
the temperature $1/\beta$ is much smaller than the typical
value of superconducting energy gap, so we can disregard
quasiparticle excitations in the superconductor.
The particle current through diffusive conductor is
given by the drop of the distribution function at its ends,
the particle current via the interface is given by the corresponding
block of Eq. \ref{main}. This yields
\begin{eqnarray}
\bm{i}_{B}^{(p)}+\bm{i}_{D}^{(p)}=0 \\
\bm{i}_{B}^{(p)}
=f_t\frac{2e^{2}}{\rm{h} }
\sum_{m} T^*(\gamma,\epsilon,T_m), \ \
\bm{i}_{D}^{(p)}
=(f_t-f_0)/R_D \nonumber
\end{eqnarray}
$f_0$ being the symmetrized distribution function in the normal reservoir,
$f_{0} =\frac{1}{2}[\tanh (\beta(\epsilon + eV)/2)
-\tanh (\beta(\epsilon - eV)/2) ]$.
The above relation becomes especially transparent if one
regards $T^*$'s as effective transmission coefficients 
in each channel. It just shows that the full (energy-dependent)
resistance of the system is the sum of the resistance of 
diffusive metal and the interface resistance, 
the latter being influenced by proximity effect. The
degree of proximity effect is determined from Eq. 
(\ref{spectral_balance}). If we define the average over 
the angle
as 
\[
<A(\theta)> =
 \int_{-\pi/2}^{\pi/2} d\theta \cos\theta A(\theta)
 /\int_{-\pi/2}^{\pi/2} d\theta T(\theta)\cos\theta
\]
with $T(\theta)=T_{m}$, 
both balance equations can be rewritten in a compact form. 
\begin{eqnarray}
R = R_D +R_B/<T^*(\gamma,\epsilon,T_m)>, \label{balance1}\\
\gamma = <F(\gamma,\epsilon,T_m)>R_D/R_B. \label{balance2}
\end{eqnarray}
Here $R_B$ is the interface resistance in normal state,
$R$ is the full resistance. It may depend on energy, so that
the full electric current is given by $eI_{el}= \int d\epsilon f_0(\epsilon)
/R$.

To reveal the underlying physics, we present the
concrete expressions for $F=F(\gamma,\epsilon,T_m)$ and 
$T^*=T^*(\gamma,\epsilon,T_m)$ assuming 
$\mid \epsilon \mid \ll \mid \Delta_{\pm} \mid$. 
It turns out that these expressions
are essentially different for $\Delta_{+}\Delta_{-}><0$, 
this manifesting the formation
of ABS in the latter case. For $\Delta_{+}\Delta_{-}<0$ 
(ABS channels) we have:
\begin{eqnarray}
F =\frac{-2T_m \sin \gamma}{T_m \cos \gamma - i (2-T_m) 
\epsilon/{\tilde\Delta}}
=\stackrel{\epsilon \rightarrow 0}{\longrightarrow} -2 \tan \gamma \\
T^* = \frac{ T_{m}^{2} 
(1 + \mid \cos \gamma \mid^2 + \mid \sin \gamma \mid^{2})}
{T_{m}^{2} \cos^{2} \gamma + (2-T_{m})^{2}(\epsilon/{\tilde\Delta})^{2}}
=\stackrel{\epsilon \rightarrow 0}{\longrightarrow} \frac{2}{\cos^2\gamma}
\end{eqnarray}
with $\tilde\Delta=(2 \mid \Delta_{+} \mid \mid \Delta_{-} \mid)/(\mid \Delta_{+} \mid + \mid \Delta_{-} \mid)$. 
It is somewhat counterintuitively that the zero-energy limit does not depend on
the actual transmission, giving finite currents even for insulating
interfaces. This is the signature of 
the resonance forming precisely at zero energy \cite{TK95}.
If the transmission is low, the resonance feature persists in a narrow
energy interval $\simeq T_{m} \Delta_\pm$ only.
The spectral current $F$ eventually suppresses the proximity effect.
The explanation is that ABS form a reservoir of {\it normal} electrons
within the unconventional superconductor, and $F$ can be viewed as
a connection to this normal reservoir.
The effective transmission coefficient $T^*$ at resonance
is always bigger than $2$, and is enhanced by proximity effect.
One can understand this as a {\it mupliple} Andreev reflection
induced by the corresponding ABS.

In the case of $\Delta_{+}\Delta_{-}>0$ ("conventional" channels) 
the resonance feature is absent
and energy dependence can be safely disregarded.
The expressions are identical to those of {\it conventional}
superconductor
\begin{eqnarray}
F=   \frac{2T_{m} s \cos\gamma }{ 2-T_{m} +T_{m} s \sin \gamma} \\
T^* = \frac{2T_{m}[ T_{m} + (2-T_{m})s \sin \gamma ]}
{\mid 2-T_{m} + T_{m} s \sin \gamma \mid^{2}}
\end{eqnarray}
Here $s \equiv {\rm sgn}(\Delta_{+}) ={\rm sgn}(\Delta_{-})$.
The spectral
current $F$ thus induces proximity effect of the corresponding sign $s$. 
The effective transmission $T^*$ does not exceed $2$ (which is the
limiting case of ideal Andreev reflection). 
Being compared with the transmission in the normal state, the effective
transmission is suppressed (enhanced) at $T_{m} <(>) 2/3$.
The fully developed proximity effect ($\gamma=s\pi/2$) restores the
normal transmission. \par
To summarize, the proximity effect originates from the "conventional"
channels 
and is suppressed by ABS channels. 
While the proximity
effect is present, 
it enhances transmission via ABS channels. 
It restores the effective transmission
of "conventional" channels to that in the normal state. 
The full resistance of the structure is determined
by competition of all these effects. It is essential that one can tune
the relative number of "conventional" and ABS
channels by changing the misorientation angle $\alpha$. 
As one can see
from the Fig.1, the ABS channels are there in the angle
interval $\pi/4 - |\alpha| <|\theta|<\pi/4 +|\alpha|$. If $\alpha=0$, there are
no such channels. If $\alpha = \pi/4$, there are no "conventional"
channels.
This gives no chance to proximity effect. \par
To illustrate this further, we calculate with Eqs. 
(\ref{balance1}),(\ref{balance2})
the zero-voltage resistance ($\epsilon \rightarrow 0)$
at different values of $\alpha$ as a function of $R_B/R_D$.
The angular dependence of the transmission
coefficient was assumed to be 
$T(\theta)=\cos^{2}\theta/(\cos^{2}\theta + Z )$ with barrier parameter 
$Z$.
The results are presented in Figs. 2,3.
At $R_D=0$ there is no proximity effect in "DN" and the resistances
are given by the quasi-ballistic formulas of Ref. \cite{TK95}.
The proximity effect may develop with increasing $R_D$ and decreases
the interface resistance. This gets the curves down.
The curves $(a)$ and $(a')$ correspond to $d$-wave junction at $\alpha=0$
and conventional superconductor junction, respectively.
One sees that the proximity effect is weaker
in d-wave system. 
This is due to competition of the "conventional" channels having
different signs of $\Delta_{+}$.
The ABS channels appear with increasing $\alpha$. The curve $(b)$
demonstrates interface conductance reduced slightly below
its normal state value. This manifests the enhanced transmission in ABS channels. The ABS channels quench the proximity
effect very efficiently at $\alpha > 0.02 \pi$. 
The total conductance can be approximated
by $R_D +R_{R_D=0}$, although this becomes exact only at $\alpha = \pi/4$.
Following Ref. \cite{reflec}, we regard the counterintuitive negative
sign of $(dR/dR_D)_{R_D=0}$ as a signal of importance of the proximity effect
[or reflectionless tunneling (RLT) ]. 
We evaluate this sign at different $Z$ and $\alpha$. (Fig. 3)
The sign of $dR/dR_D$ is negative for junctions of low
transmissivity in a relatively narrow range of $\alpha$.

In conclusion, we have extended the circuit theory of 
superconductivity to include unconventional superconductor
junctions. We have derived a general relation for
matrix current to/from unconventional superconductor.
An elaborated example demonstrates the interplay 
of ABS and proximity effect in a $d$-wave junction. 
The theory presented will facilitate the analysis
of more complicated mesoscopic systems that include
unconventional superconductors.  \par
%
The authors appreciate useful and fruitful discussions
with  J. Inoue, H. Itoh and  G.W.E. Bauer.
%
%
%
\begin{figure}[hob]
\begin{center}
\includegraphics[width=5cm,clip]{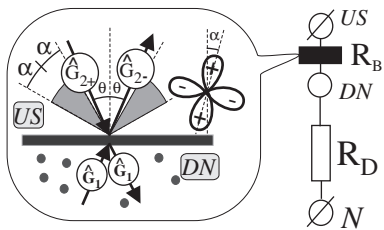}
\end{center}
\vskip -4mm
\caption{
The unconventional superconductor junctions (black box) 
can be incorporated into circuit theory by means of 
the matrix current relation (\ref{main}). This relation
accounts for anisotropic features of the US, as sketched
for a $d$-wave superconductor.
}
\label{fig:01}
\end{figure}

\begin{figure}[htb]
\vspace{0.5cm}
\begin{center}
\includegraphics[width=5cm,clip]{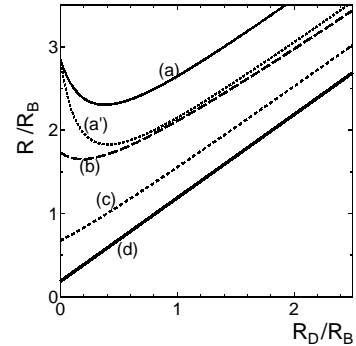}
\caption{
Full resistance of the circuit versus $R_D$
for various $\alpha$.
(a): $\alpha=0$,
(b): $\alpha=0.01\pi$,
(c): $\alpha=0.05\pi$,
and 
(d): $\alpha=0.25\pi$. $Z$=1. 
The curve (a') presents the same dependence
for a conventional superconductor.
Similar results for $\alpha=0$  and $\alpha=\pi/4$ were 
obtained in [14] by numerical 
simulations. 
\label{fig:02}}
\end{center}
\end{figure}

\begin{figure}[htb]
\begin{center}
\includegraphics[width=5cm,clip]{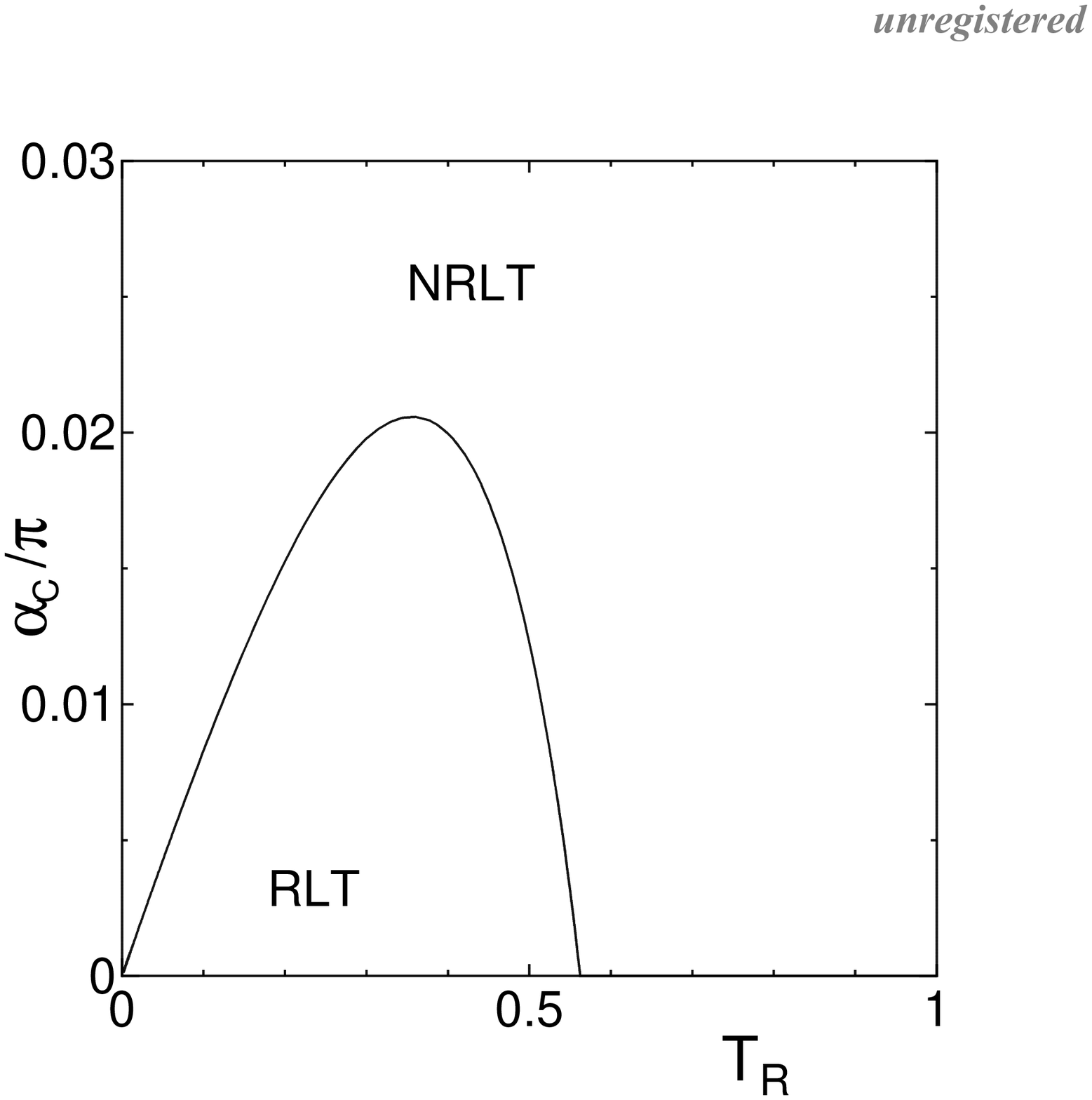}
\end{center}
\vskip -4mm
\caption{ "RLT" ("NRLT") marks the region 
where $dR/dR_{D}<0$ ($dR/dR_{D}>0$)  at 
$R_{D}=0$. 
$T_R$ is the average transmissivity of the junction 
in the normal state, $T_R \equiv 
1 - \frac{Z}{2\sqrt{Z+1}}\ln [\frac{\sqrt{Z+1}+1}{\sqrt{Z+1}-1}]$
for the model in use. 
}
\label{fig:04}
\end{figure}

\end{document}